\documentclass[runningheads]{llncs} 
\usepackage[T1]{fontenc} 
\usepackage{graphicx} 

\usepackage{float}
\begin{document}
\title{Research on the Application of Spark Streaming Real-Time Data Analysis System and large language model Intelligent Agents}

\author{Jialin Wang\inst{1}  \and
Zhihua Duan\inst{2 }   }
 
\institute{
Executive Vice President,Ferret Relationship Intelligence\\Burlingame, CA 94010, USA \\
\email{jialinwangspace@gmail.com}\\
\url{https://www.linkedin.com/in/starspacenlp/} 
\and
Intelligent Cloud Network Monitoring Department \\
China Telecom Shanghai Company,Shanghai, China\\
\email{duanzh.sh@chinatelecom.cn}\\
}

\maketitle              % typeset the header of the contribution
\begin{abstract}
This study explores the integration of Agent AI with LangGraph to enhance real-time data analysis systems in big data environments. The proposed framework overcomes limitations of static workflows, inefficient stateful computations, and lack of human intervention by leveraging LangGraph’s graph-based workflow construction and dynamic decision-making capabilities. LangGraph allows large language models (LLMs) to dynamically determine control flows, invoke tools, and assess the necessity of further actions, improving flexibility and efficiency.

The system architecture incorporates Apache Spark Streaming, Kafka, and LangGraph to create a high-performance sentiment analysis system. LangGraph’s capabilities include precise state management, dynamic workflow construction, and robust memory checkpointing, enabling seamless multi-turn interactions and context retention. Human-in-the-loop mechanisms are integrated to refine sentiment analysis, particularly in ambiguous or high-stakes scenarios, ensuring greater reliability and contextual relevance.

Key features such as real-time state streaming, debugging via LangGraph Studio, and efficient handling of large-scale data streams make this framework ideal for adaptive decision-making. Experimental results confirm the system’s ability to classify inquiries, detect sentiment trends, and escalate complex issues for manual review, demonstrating a synergistic blend of LLM capabilities and human oversight.

This work presents a scalable, adaptable, and reliable solution for real-time sentiment analysis and decision-making, advancing the use of Agent AI and LangGraph in big data applications.

\keywords{Large Language Model \and Agent \and Langchain \and ChatGPT  \and ERNIE-4  \and GLM-4 \and Qwen2.5  \and Big Data  \and Spark Streaming \and Real-time data analysis system \and Sentiment Analysis .}
\end{abstract}
 
\section{Introduction}
In the internet era, there are higher requirements for the effectiveness and granularity of data analysis. At present, social applications, e-commerce and new media applications all have generated massive data. In order to mine and analyze the information value in data stream in real time, it is necessary to design a real-time data processing system for big data analysis. 

The frameworks and technologies for big data processing have been developed continuously from Map Reduce, a parallel distributed processing framework based on Hadoop platform, to Storm streaming processing and Spark ecosystem, which provide distributed big data processing methods and programming interfaces, which have a great effect on the development of massive data analysis programs.

Radix calculation is often used to implement data analysis, that is to say, calculating the number of different elements in the data \cite{1}.For example, to count the number of independent IP addresses within 5 minutes of a website. We can use an accurate radix calculation algorithm to work out this problem, but it needs to cache all IP addresses within 5 minutes, which consumes too much resources. While the radix estimation algorithm is an algorithm based on probabilistic statistics theory to estimate the radix of different elements in a given data set, which improves the computational efficiency by sacrificing certain data accuracy\cite{2}.

The main contents of this paper are as follows: Section 2 designs a real-time data acquisition and analysis system based on Spark Streaming; Section 3 compares and analyses the accurate calculation methods and estimation methods of radix calculation for stateful computing operations, and verifies that the radix estimation method based on HyperLog+is more suitable for real-time statistics of radix estimation for big data; Section 4 makes a summary about this paper. 

With the rapid development of large model technology, real-time data analysis based on intelligent agents is becoming increasingly important. This paper will delve into the construction of an efficient real-time data analysis system through advanced technologies such as Apache Spark, Spark Streaming, Large Language Model  intelligent agents, and Apache Kafka. Apache Spark, as a powerful distributed computing framework, not only handles large-scale datasets but also supports real-time data stream processing. With real-time analysis based on large model intelligent agents, it can provide dynamic decision support. 

\section{Real-time Data Acquisition and Analysis System}
Real-time data acquisition and analysis system not only needs to meet the requirements of concurrency, but also needs to ensure real-time data processing and certain data disaster tolerance guarantees. Based on open source system, this paper designs a real-time data acquisition and analysis system including data acquisition service, data queue service, and data analysis service
[3,4]. The overall architecture includes data acquisition client-side, data acquisition server OpenResty, Kafka (Distributed Publish/Read Message System) Cluster and Data Analysis and Computing Program based on Spark Streaming-based

\subsection{Client-Side of Data Acquisition and its Format Definition} 

In order to make it easier for third-party applications to implement data integration and uniform data processing, the Client-Side of Data Acquisition (Open Resty) is designed, and the format of the collected data in the client is defined. According to the subject of the event, the category of the event, the attributes involved in the event, the time of the event, the location of the
event and the result of the event, the log information is defined, which mainly includes several domains: application identification domain, device information domain, user identification domain, action event domain, action object domain, action time domain, action geography domain and action result domain, as shown in Table 1.

\begin{table}
\caption{Definition of Log Information Domain.}\label{tab1}
\begin{tabular}{|p{0.4\linewidth}|p{0.6\linewidth}|}
\hline
Log Information Domain & Description\\
\hline
Application Identification Domain & Identity, version and type of application are defined. \\
\hline
Equipment Information Domain & Operating system, resolution, model,  resolution and useragent of the device are defined. \\
\hline
User ID Domain & Definition of device identification and  user identification\\
\hline
Action Event Domain & Common browsing events, abnormal events, playing  related events, searching, commenting, sharing and other  events are defined. Custom events are also supported. \\
\hline
Action Object Domain & Business attributes related to the corresponding  event domain are defined, such as node recognition, URL address,  stream address, playback time point, etc. \\
\hline
Action Time Domain & The start and end times of events are defined\\
\hline
Action Geography Domain & The latitude and longitude of event occurrence,  network type, IP address and so on are defined. \\
\hline
Action Result Domain & Define the results of events\\
\hline
\end{tabular}
\end{table}

\subsection{Real-time Data Acquisition Service}  
Nginx is a high-performance HTTP server, which has the advantages of low memory occupation and high stability compared with Apache, Lighthttpd, and other HTTP servers. OpenResty is a high-performance Web platform based on Nginx server and Lua interpreter, which makes full use of Nginx's non-blocking I/O model and uses Lua scripting language to call various C and Lua modules supported by Nginx. It not only makes HTTP client connection requests but also provides consistent, high-performance responses to remote backends such as MySQL, Redis, and Kafka.

Based on OpenResty's real-time data acquisition service, the format of the report log is checked to avoid abnormal log entry and polluting the system through Lua module based on the defined log record format. At the same time, the user agent information and IP address information in the log are supplemented and processed. Through the Kafka Lua producer client, the processed log is released to different Kafka partitions for data collection. 

\subsection{Real-time Data Cache Queue}
Kafka is an easy-to-extend topic-based publish/subscribe message queuing system, which mainly consists of Producer, Broker, and Consumer \cite{5}. Producer is responsible for collecting and sending messages to Broker. Broker receives messages from Producer and persists them. Consumer is the user of messages and gets messages from Broker. Kafka, as a buffer of data aggregation, is the core of the whole data architecture, which can provide data for multiple consumers and enable real-time data to serve multiple scenarios and services. 

The Kafka cluster decouples the real-time data flow and achieves asynchronous processing between the producer and the consumer of the data. As a Producer of real-time data, the data acquisition server, OpenResty, publishes to Kafka cluster topics based on event types. Spark Streaming, as a Consumer, reads data from Kafka cluster, forming a real-time data processing pipeline.

\subsection{Real-time Data Analysis System Based on Spark Streaming}
Spark Streaming is a frame used for real-time calculation in Spark big data analysis system. The Spark Streaming splits real-time data streams into DStream, Discretized Stream. As a basic abstraction in Spark Streaming, internal DStream maintains a set of Resilient Distributed Datasets with discrete time axis as the key \cite{6}. These RDD sequences respectively represent datasets in different time periods, and various operations on DStream will eventually be mapped to internal RDD, so as to achieve a seamless connection with Spark.

\subsection{Real-time Data Analysis and Classification}
Data analysis based on Spark Streaming usually includes two kinds of calculation: stateless calculation and stateful calculation. For example, when calculating the number of page views per 5 minutes and the number of independent IP, if the batch processing interval of Spark Streaming is 5 minutes, the number of PVs per day can be directly added up and calculated; while the number of independent IP needs to save all the IP numbers in 5 minutes before unified computing can be carried out. It will not only cost more resources to complete stateful computing, but also takes a long time to respond to the need for everyday and weekly multi-granularity statistics. Therefore, only by classifying data computing and adopting different calculation methods according to different classifications, can the analysis be more effective and real-time.

\subsection{Intelligent Agent Spark Streaming Analysis}
As shown in Figure 1, this study implements a Spark Streaming real-time analysis system architecture based on large language model agents, achieving research on sentiment analysis of user review data.

\begin{figure}[H]
\includegraphics[width=\textwidth]{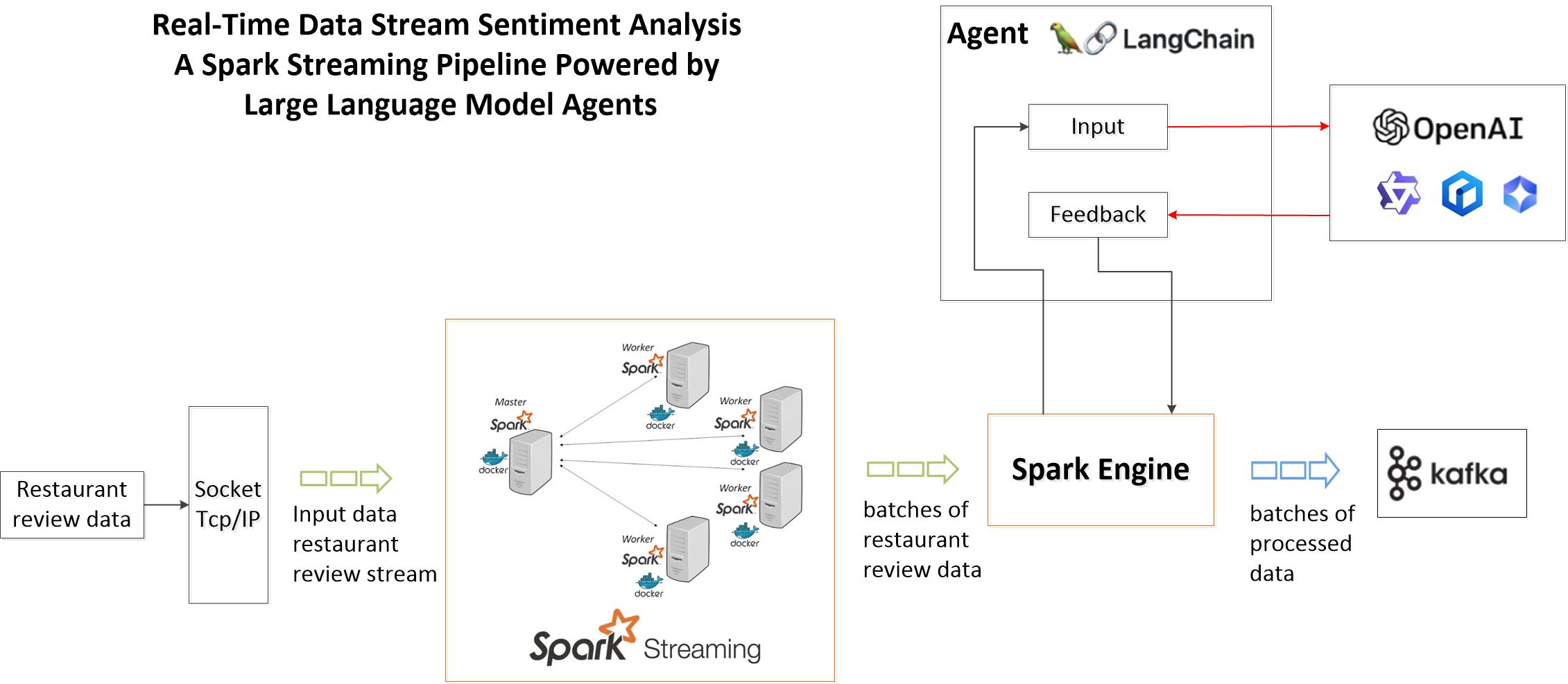}
\caption{Spark Streaming and Large Model Intelligent Agent Architecture Diagram.} \label{fig1}
\end{figure} 
This system primarily employs the following key technologies to achieve efficient data processing and analysis:

1. Sockets: Serving as the initial data access point, it facilitates network communication through the TCP/IP protocol for data transmission between clients and servers. In this study, restaurant review datasets and Yelp datasets are used as experimental data.

2. Spark Streaming: Apache Spark is an efficient distributed computing framework that offers fast and versatile data processing capabilities. The Spark Streaming module is specifically designed for handling real-time data streams. This experiment configures and deploys Spark-related services and containers using Docker, setting up a cluster architecture with a Spark master and workers.

3. Spark Engine: Spark Streaming receives real-time data streams and processes them in batches. The Spark engine is responsible for handling these batched data and generating result streams. The Spark engine also interacts with large model agents.

4. Large Model Agents: Providing sentiment analysis capabilities, these agents are based on large language models, enhancing the data processing and analysis capabilities for sentiment analysis, allowing the agents to more accurately understand and process natural language data. Models such as GPT-4, Qwen 2.5, ERNIE 4.0, and GLM-E 4 can be utilized.

5. Kafka: Apache Kafka is a distributed streaming platform that, when integrated with Spark Streaming, can efficiently handle real-time data streams. The processed data is output to Elasticsearch, which is responsible for storing and indexing the processed data, enabling rapid retrieval and visualization with Elasticsearch and Kibana.

This system constitutes a powerful real-time data processing and analysis application, capable of handling large-scale datasets and providing real-time sentiment analysis capabilities.

\section{Accurate calculation algorithm and approximate estimation algorithm}
Radix calculation method is a method to determine the number of different elements in a data stream, which consists of exact calculation algorithm and approximate estimation algorithm. Accurate calculation algorithm can usually be easily calculated by linear space complexity O(N) algorithm, but it often requires a lot of memory as well as long time, which makes itself often unable to meet the needs to process massive data. Therefore, the radix estimation method for mass data with less resources comes into being. Generally speaking, radix estimation algorithms are mainly Linear Counting, LogLog Counting, HyperLogLog Counting \cite{2} and Adaptive Counting \cite{1}. HyperLogLog++ \cite{7} is an algorithm based on the HyperLogLog Counting algorithm with low error. This paper mainly estimates the number of independent IP per day based on HyperLog+ algorithm.

\subsection{Accurate Computing Algorithms}
The accurate radix calculation method, which is based on SparkStreaming, calculates the number of IP that appears for the first time every day in each batch according to the historical status, and then adds the number of IP in each batch, which refers to the number of independent IP every day. The specific process and pseudocode are shown in figure 2 and figure 3 respectively:

\begin{figure}[H]
\includegraphics[width=\textwidth]{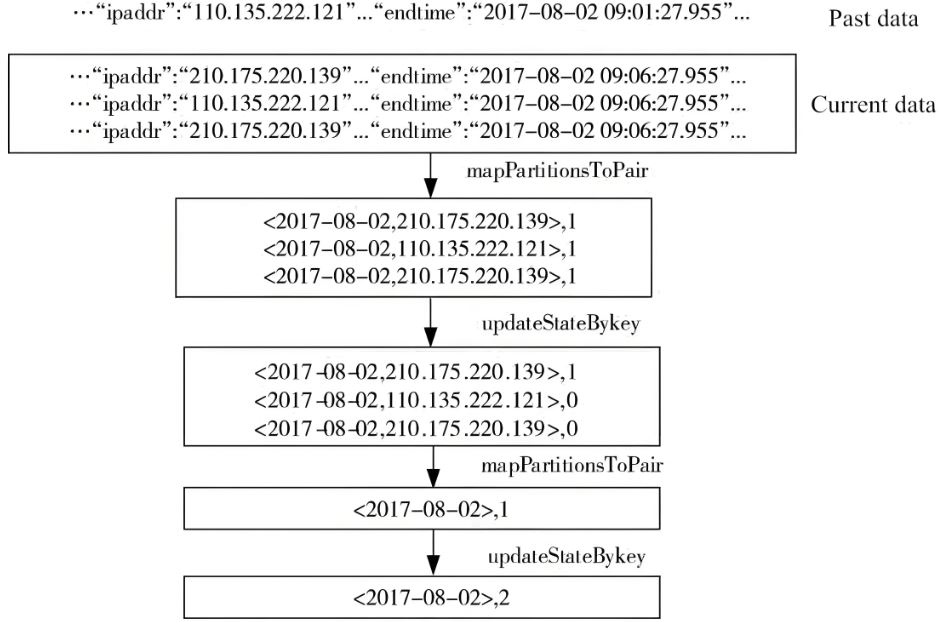}
\caption{Accurate calculation method flow for calculating the number of independent IPs per day based on SparkStreaming statistics.} \label{fig2}
\end{figure}
\begin{figure}[H]
\includegraphics[width=\textwidth]{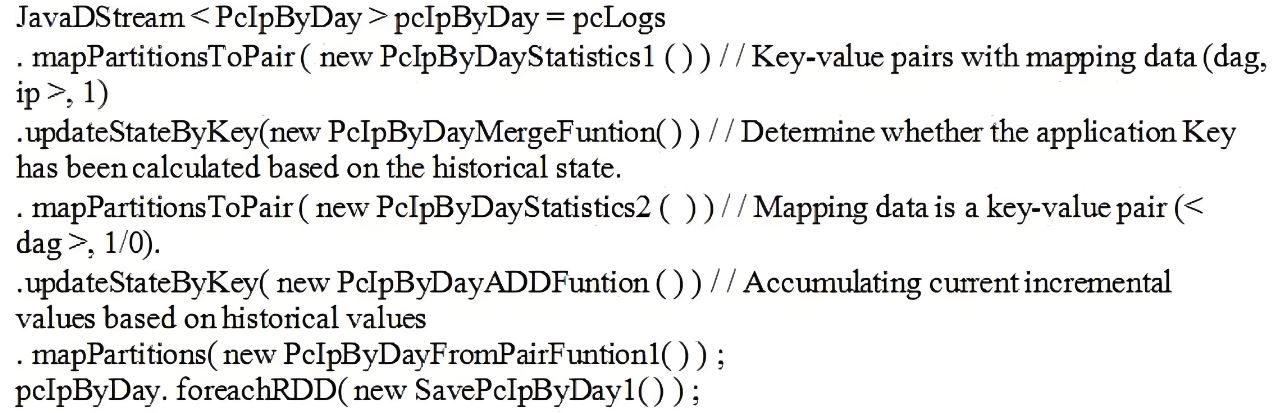}
\caption{ Pseudocode based on SparkStreaming.} \label{fig3}
\end{figure}

\subsection{Radix Estimation Method}
Spark Streaming-based Radix Estimation Method saves a HyperLogLog++ object to the history state per day  and the IP of each batch will be added to the corresponding HyperLogLog++ object. When calculating, call HyperLogLog++ objects to obtain the number of daily independent IP  and the specific process and pseudocode are shown in figure 4 and figure 5 respectively:
\begin{figure}[H]
\includegraphics[width=\textwidth]{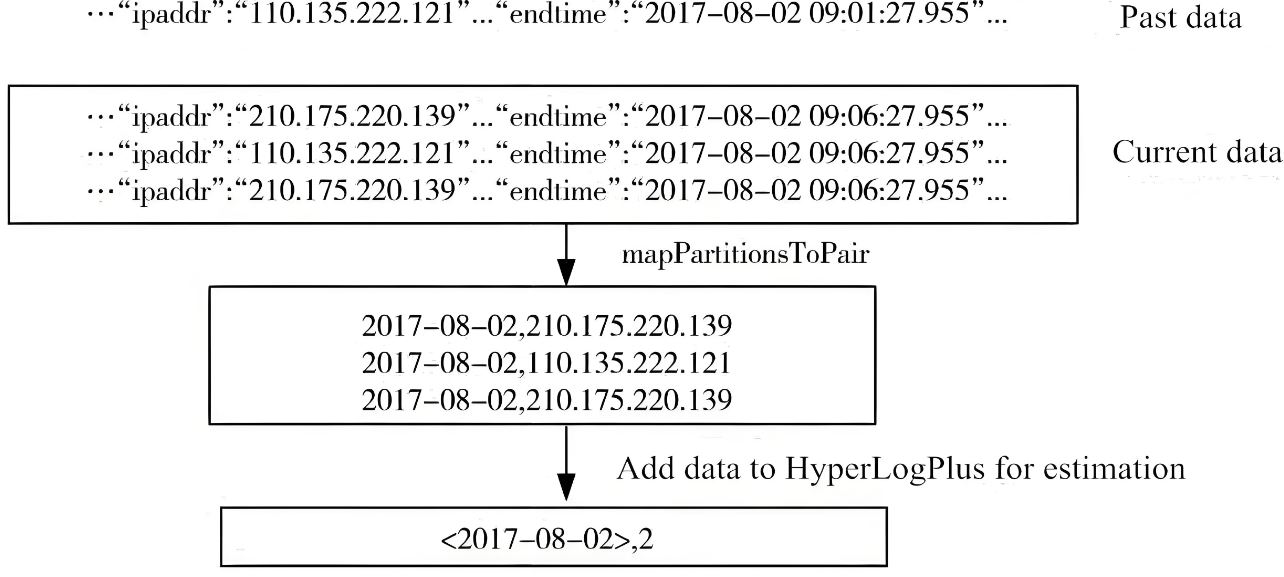}
\caption{HyperLogLog++ Estimation Method for Statistics of Daily Independent IP Number.} \label{fig4}
\end{figure}

\begin{figure}[H]
\includegraphics[width=\textwidth]{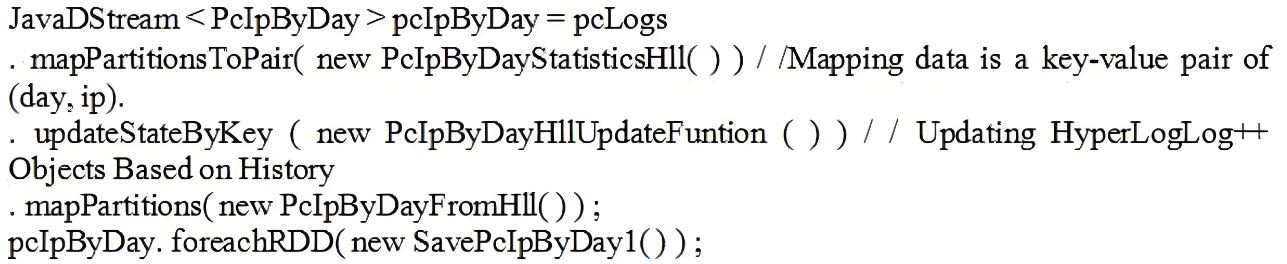}
\caption{ Pseudo-code based on HyperLogLog++.} \label{fig5}
\end{figure}

\section{Analysis of Spark Machine Learning Algorithms}

The machine learning library of Apache Spark (MLlib) is a key component of the Spark ecosystem, covering various machine learning tasks such as classification, regression, clustering, and collaborative filtering. Leveraging the distributed computing architecture of Apache Spark, MLlib can scale to more computing nodes with the increase of data volume, processing larger datasets. This not only enhances the capability of data processing but also accelerates the model training and inference process on large datasets.

Spark MLlib provides a comprehensive library of algorithms, supporting a variety of machine learning algorithms, including linear regression, logistic regression, decision trees, random forests, and k-means clustering, among others. These algorithms can adapt to different business scenarios and data characteristics, meeting a diverse range of analytical needs. Additionally, MLlib supports key aspects of feature engineering, enabling the construction of complex pipelines involving multiple machine learning techniques, such as feature extraction, feature transformation, and feature selection.

Spark MLlib can be seamlessly integrated with Spark Streaming, enabling machine learning analysis on real-time data streams. This is of significant practical value for business scenarios requiring real-time feedback. It not only improves the efficiency of data processing but also provides strong support for real-time data analysis.

\section{Analysis of experimental results}
Three experiments were carried out according to three conditions of 3000, 5000, and 10000 log records being produced in each batch. Each experiment was run 12 batches, totaling 36 batches. In each batch, 1/2 of the number of IP is the same. In the experiment, the batch processing interval of Spark Streaming was 5 minutes, i.e. 3000, 5000, or 10000 log records were generated every 5 minutes. The size of each log record was 1,296 bytes, and the IP address in the log is 15 bytes.

Accurate Radix calculation method and Radix estimation method based on HyperLogLog++ are run on the log respectively. In the process, the program is deployed and submitted in local mode, and the number of independent IP, the processing time of each batch, and the size of the file size under the checkpoint path are counted. 

The calculation results of the error percentage of the number of independent IP counted in each batch in the estimation counting method are shown in Table 2. From Table 2, we can see that compared with the accurate calculation method, the error rate of the estimation method is less than 1.5\%. For the number of daily independent IP in each batch, the error can be neglected. Moreover, from the perspective of the number of log records processed in batches, the more the number of logs, the fewer the relative error is.
\begin{figure}
\includegraphics[width=\textwidth]{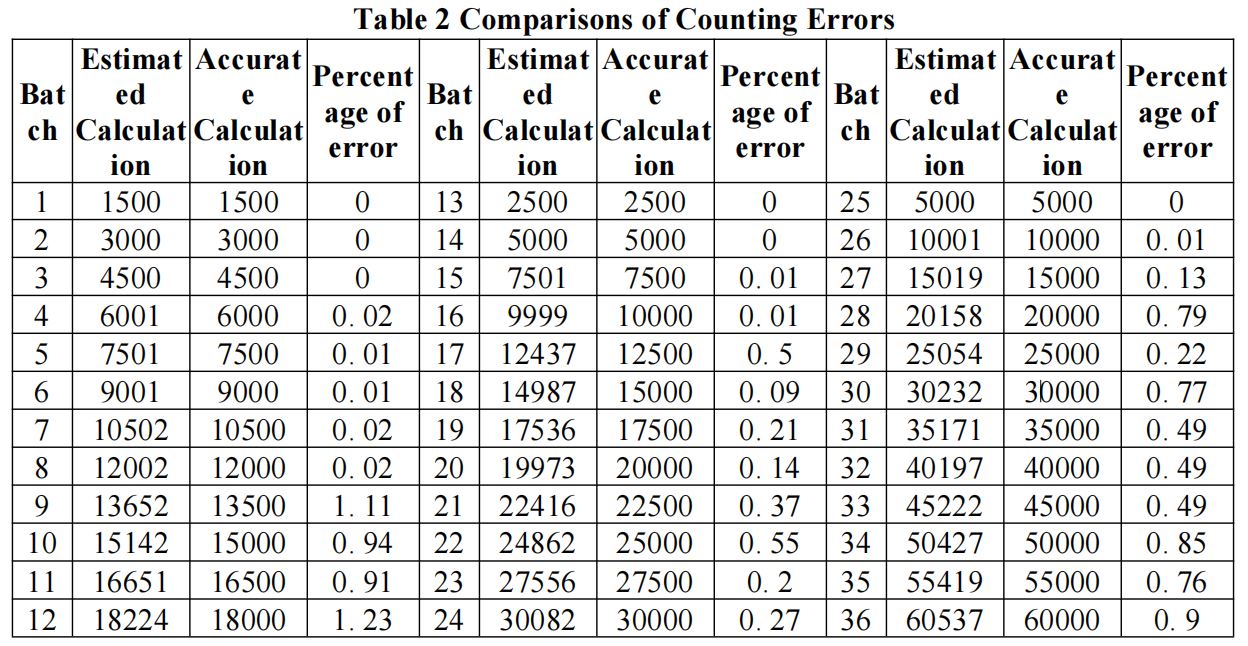}
 
\end{figure}

The statistical results of processing time for each batch are shown in Fig 6. From Fig 6, we can see that with the same quantities of logs, the average processing time for each batch of the accurate calculation method is twice as long as that of the radix estimation method based on HyperLogLog++. Furthermore, as the number of log records increases, the average processing time per batch of the accurate calculation method increases more than that of the radix estimation method based on HyperLogLog++. It can be seen that in the batch of the third experiment, the batch processing time of the accurate calculation method is longer than the batch processing interval of Spark Streaming, resulting in a larger scheduling delay.

\begin{figure}
\centering
\includegraphics[width=0.8\textwidth]{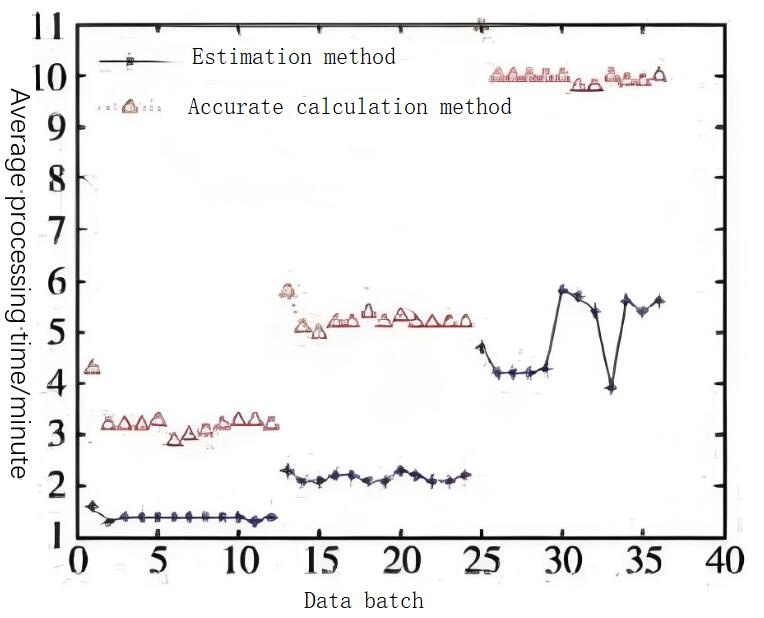}
\caption{Batch processing time statistics.} \label{fig6} 
\end{figure}
Figure 7 shows the change of checkpoint occupancy with batches. From Figure 7, it can be seen that the storage space usage based on HyperLogLog++ tends to be stable with small fluctuation, and the absolute amount of occupancy space is much lower than the accurate calculation method, whose occupancy space increases dramatically with the increasing number of log records.

\begin{figure}
\centering
\includegraphics[width=0.8\textwidth]{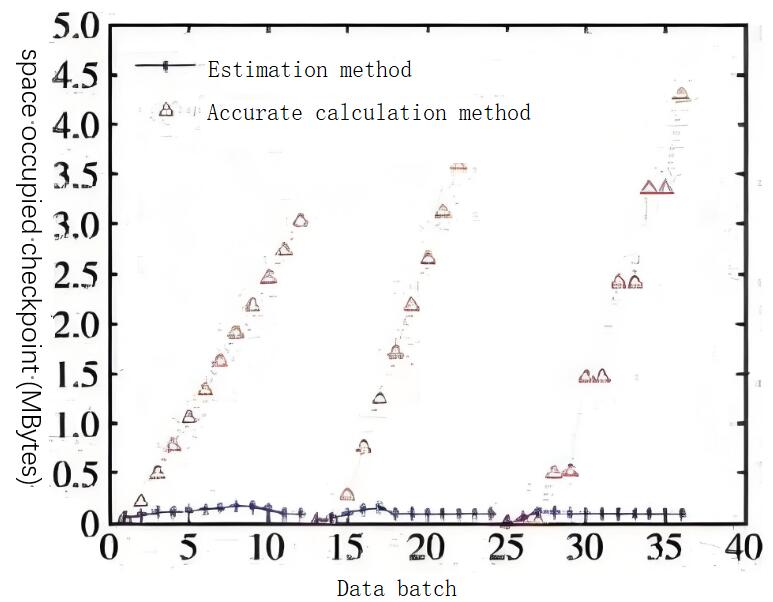}
\caption{Checkpoint occupancy statistics.} \label{fig7} 
\end{figure}

According to the above analysis, it can see that the HyperLog+ based radix estimation method has obvious advantages over the accurate calculation method in processing time and checkpoint occupancy space, with the error rate of below 1.5\%, which can be basically neglected. Therefore, the HyperLogLog+ radix estimation method is more suitable for real-time statistics of radix estimation of big data.

\section{Discussion}
\subsection{Limitations of Real-Time Systems}
The real-time system proposed in the paper has limitations in dynamic workflow management, stateful computing, and human intervention, but these can be improved by using LangGraph to implement workflows with conditions and loops, state saving, and mechanisms for human review and feedback, thereby enhancing system efficiency.

1. Dynamic Workflow Management
Current Limitation: The paper lacks an explicit mechanism for handling dynamic and adaptive workflows. It discusses static pipelines for data processing, sentiment analysis, and radix estimation, but does not address flexibility in changing tasks or branching logic in workflows.

Improvement Using LangGraph:Cycles and Branching: Use LangGraph to implement workflows with conditionals and loops. For example:
If data is incomplete, trigger preprocessing workflows.
If errors occur in sentiment analysis, retry or escalate to a human operator.
Controllability: Enable fine-grained control over the execution flow, ensuring robust handling of diverse data conditions or model requirements.
Rationale: This improves system resilience, allowing it to handle edge cases and dynamic data streams efficiently.

2. Persistence for Stateful Computations
Current Limitation: The paper's architecture does not address state persistence explicitly. Stateful computations like IP tracking rely on in-memory or ephemeral mechanisms, which could lead to inefficiencies or data loss in distributed environments.

Improvement Using LangGraph:Built-in Persistence: Implement state-saving after each step of the graph. For example:
Save intermediate results of HyperLogLog++ estimations.
Persist checkpoint data in case of failures, enabling seamless recovery.
Error Recovery and Time Travel: Add the ability to rewind workflows for debugging or replaying historical data.
Rationale: Enhances reliability, particularly for long-running processes, while reducing the risk of data loss.

3. Human-in-the-Loop for Enhanced Sentiment Analysis
Current Limitation: Sentiment analysis in the paper relies entirely on LLMs without any mechanism for human oversight or correction, which may lead to inaccuracies in edge cases or domain-specific contexts.

Improvement Using LangGraph:Human-in-the-Loop: Incorporate LangGraph’s feature to pause workflows for human review and input. For example:
When sentiment analysis confidence is low or results are ambiguous, notify a human operator for validation or refinement.
Use human feedback to improve model performance iteratively.
Rationale: Improves accuracy and contextual relevance of sentiment analysis, especially in high-stakes applications.

\subsection{Improvement Solutions Based on LangGraph
}
As shown in Figure 8.This study employs LangGraph technology to optimize and upgrade a big data analysis system. LangGraph, as an advanced tool for constructing large language model workflows, integrates large language models with graph-based workflows to implement an emotional analysis agent system. This system is not only capable of efficiently categorizing customer inquiries but also provides manual responses or escalates issues to higher-level processing procedures when it detects negative customer sentiment.

The key technical features of the system include:

1. State Management  : Precisely define and manage the state of customer interactions.

2. Graph Construction  : Utilize StateGraph to build workflows, designing nodes, edges, and conditional edges to represent complex support processes.

3. Memory Checkpoint Settings  : Save state and context information in multi-turn conversations to achieve coherence in dialogue and persistence of context.

4. Human in the Loop  : In cases requiring human intervention, the system can facilitate manual intervention, allowing for the setting and updating of responses to enhance the quality of responses.

By applying these technologies, the LangGraph system can handle complex customer interactions more flexibly while ensuring that manual intervention can be introduced in a timely manner when necessary, thereby improving the efficiency and quality of customer service.

\begin{figure}[H]
\centering
\includegraphics[width=0.8\textwidth]{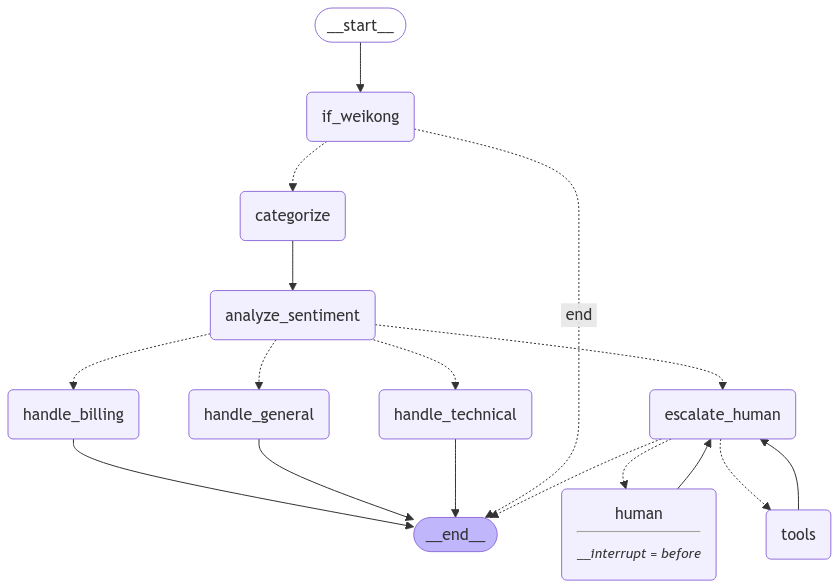}
\caption{Sentiment Analysis Based on LangGraph.} \label{fig8} 
\end{figure}

\section{Conclusion}
This study demonstrates how Agent AI, empowered by LangGraph, transforms real-time data streaming systems by introducing dynamic control flow, persistent state management, and human-in-the-loop workflows. LangGraph’s ability to implement cycles and branching in workflows allows streaming systems to adapt to complex and evolving data processing requirements, ensuring reliable and efficient performance. By integrating LangGraph’s persistence features, the system supports seamless error recovery, memory retention, and iterative processing essential for high-stakes applications.

In conjunction with Apache Spark Streaming and Kafka, Agent AI leverages LangGraph’s streaming support and debugging tools to enhance sentiment analysis workflows. This integration enables real-time decision-making, adaptive task execution, and the refinement of insights through human collaboration. Experimental validations highlight the system’s capacity to dynamically manage workflows, process large-scale data efficiently, and deliver actionable intelligence.

By blending LangGraph’s advanced capabilities with Spark Streaming, this framework establishes a robust foundation for scalable, intelligent, and adaptive streaming systems, advancing the application of Agent AI in big data environments.

\end{document}